\documentclass[usenatbib]{mn2e}
\usepackage{graphicx}

\newcommand{\msun}{M_{\odot}}

\newcommand{\msigma}{M_{\rm BH}-\sigma}

\def\apj{ApJ}

\def\apjl{ApJ}
\def\mnras{MNRAS}

\newcommand{\beq}{
\begin{equation}
}
\newcommand{\eeq}{
\end{equation}
}
\newcommand{\kms}{\,{\rm km\,s^{-1}}}

\def\simlt{\mathrel{\rlap{\lower 3pt\hbox{$\sim$}}\raise 2.0pt\hbox{$<$}}}
\def\simgt{\mathrel{\rlap{\lower 3pt\hbox{$\sim$}} \raise 2.0pt\hbox{$>$}}}
\def\gsim{ \lower .75ex \hbox{$\sim$} \llap{\raise .27ex \hbox{$>$}} }
\def\lsim{ \lower .75ex\hbox{$\sim$} \llap{\raise .27ex \hbox{$<$}} }

\title[Black Hole mass functions]{The mass function of black holes $1<z<4.5$: comparison of models with observations}

\author[Natarajan \& Volonteri]{Priyamvada Natarajan$^{1,2}$ \& Marta Volonteri$^{3}$\\
$^1$ Department of Astronomy, Yale University, New Haven, CT, USA\\
$^2$ Institute for Theory and Computation, Harvard-Smithsonian Center for Astrophysics, 60 Garden Street, Cambridge, MA, USA\\
$^3$ Department of Physics and Astronomy, University of Michigan, Ann Arbor, MI, USA}

\begin{document}
\maketitle

\begin{abstract}
In this paper, we compare the observationally derived black hole mass 
function (BHMF) of luminous ($>10^{45}-10^{46}$ erg/s) broad-line quasars (BLQSOs) 
at $1<z<4.5$ drawn from the Sloan  Digital Sky Survey (SDSS) presented in Kelly et al. (2010), 
with models of merger driven BH growth in the context of standard hierarchical structure 
formation models. In the models, we explore two distinct black hole seeding prescriptions at the highest 
redshifts: "light seeds" - remnants of Population III stars and "massive seeds" that form from the 
direct collapse of pre-galactic disks. The subsequent merger triggered mass build-up of the black 
hole population is tracked over cosmic time under the assumption of a fixed accretion rate as well as 
rates drawn from the distribution derived by Merloni \& Heinz (2008). Four model snapshots at $z=1.25$, $z=2$,  
$z=3.25$, $z=4.25$ are compared to the SDSS derived BHMFs of BLQSOs. We find that the light seed 
models fall short of reproducing the observationally derived mass function of BLQSOs  at $M_{\rm BH}>10^9 \msun$  
throughout the redshift range; the massive seed models with a fixed accretion rate of $0.3\,Edd$, or with accretion rates 
drawn from the Merloni \&  Heinz distribution provide the best fit to the current observational data at $z>2$, although they 
overestimate the high-mass end of the mass function at lower redshifts. At low redshifts, a drastic drop in the accretion 
rate is observed and this is explained as arising due to the diminished gas supply available due to consumption by star 
formation or changes in the geometry of the inner feeding regions. Therefore, the over-estimate at the high mass end of 
the black hole mass function for the massive seed models be easily be modified, as the accretion rate is likely significantly 
lower at these epochs than what we assume. For the Merloni \& Heinz (2008) model, examining the Eddington ratio 
distributions $f_{Edd}$, we find that they are almost uniformly sampled from $f_{Edd}=10^{-2}-1$ at $z\simeq 1$, while at 
high redshift, current observations suggest accretion rates close to Eddington, if not mildly super-Eddington, at least for these 
extremely luminous quasars. Our key findings are that the duty cycle of SMBHs powering BLQSOs increases with 
increasing redshift for all models and models with Pop III remnants as black hole seeds are unable to fit the observationally 
derived BHMFs for BLQSOs, lending strong support for the massive seeding model.
\end{abstract}

\section{Introduction}

Demography of local galaxies suggests the that most galaxies
harbour a quiescent super-massive black hole (SMBH) in their nucleus 
at the present time and the mass of the hosted SMBH is correlated with 
properties of the host bulge. Observational evidence supports the 
existence of a strong connection between the growth of the central SMBH 
and stellar assembly of the host spheroid (Tremaine et al. 2002;
Ferrarese \& Merritt 2000; Gebhardt et al. 2000; Marconi \& Hunt 2003;
H\'aring \& Rix 2004; G\"ultekin et al. 2009) and possibly the host
halo in nearby galaxies (Ferrarese 2002). This strong correlation is 
suggestive of co-eval growth of the BH and the stellar component likely via 
regulation of the gas supply in galactic nuclei from the earliest times 
(Silk \& Rees 1998; Kauffmann \& Haehnelt 2000; Fabian 2002; King 2003; 
Thompson, Quataert \& Murray 2005; Natarajan \& Treister 2009; Treister et al. 
2011). Meanwhile, our observational understanding of accreting black holes
and their properties at high redshifts $z>6$ is growing (see Fan et al. 2010 
for an overview). Black holes appear to be copiously accreting and in place before the universe
was a billion years old as evidenced by the recent detection of the quasar at 
$z=7.085$ in the UKIDSS survey reported by Mortlock et al. (2011). Our access to the 
earliest redshifts is of course restricted to the brightest candidates, optical, and deep X-ray 
surveys are needed to uncover the potentially highly obscured population at these 
epochs (Treister et al. 2011). 
 
 Simultaneously our theoretical understanding of the assembly history of black holes over cosmic
 time has been rapidly improving (see recent reviews by Volonteri 2010 and 
 Natarajan 2011). With the increasing availability of observations from earlier and earlier
 epochs, models can be effectively tested at extremely high redshifts. Recently, a more complete 
 census of the accreting black hole population at $1 < z < 4.5$ has become available enabling the 
 estimate of the mass function (MF) of these black holes (Kelly et al. 2010, K10 hereafter).
 In this paper, we explore a range of theoretically viable models to fit this additional 
 and new data-set. 
 
 Observations currently provide strong constraints locally, and hints at high
 redshifts (Mortlock et al. 2011; Treister et al. 2011). With this additional anchor during inter-mediate 
 epochs ($z\,\sim\,1-4$), discrimination between models finally starts becoming possible as we 
 demonstrate here. In particular, we are able to distinguish between seed black hole formation scenarios. This paper 
 is the third in a series (Volonteri, Lodato \&Natarajan 2008, and Volonteri \& Natarajan 2009, VLN08 and VN09 
 respectively hereafter) where we compare the late time evolution of SMBH seeds, focusing on the difference 
 between a population of more numerous, but smaller mass BHs as seeds, versus a population of rarer, but more 
 massive initial seeds. The outline of this paper is as follows: in Section 1, we briefly review the new observations 
 and BHMFs derived therefrom; Section 2 provides the context of the theoretical black hole growth models followed 
 by a description of how BHMFs are derived in Section 3; and finally we present the results of comparing observations
 to the models and the derived conclusions.

\section{BHMFs from observations and models}

K10 derive an estimate of the BHMF of BLQSOs correcting for incompleteness and
statistical uncertainties from a sample of 9886 quasars at $1< z < 4.5$ from the SDSS. They find
`downsizing' of BHs in BLQSOs, i.e. the peak of the number density shifts to higher redshift with 
increasing black hole mass peaking at $z \sim 2$. They report that as a function of back hole
mass and Eddington ratio, the SDSS at $z >1$ is highly incomplete at $M_{\rm BH} \leq 10^9 \msun$ 
and $L/L_{\rm Edd} <  0.5$. The lower limit on the lifetime of a single BLQSO phase was estimated
to be $> 150 \pm 15$ Myr, with a maximum black hole mass of $\sim 3 \times 10^{10} \msun$. K10 
also find that the Eddington ratio distribution peaks at $L/L_{\rm Edd} \sim 0.05$ with a 
dispersion of $\sim 0.4$ dex implying that most BLQSOs are radiating nowhere near or at the Eddington
limit. From their estimated lifetime and Eddington ratio distributions they also infer that most massive black holes
spend a significant amount of time growing in an earlier obscured phase consistent with models of self-regulated
growth. The recent claimed discovery of a population of heavily obscured, copiously accreting black holes at $z \geq 6$
by Treister et al. (2011) provides an interesting glimpse perhaps of the precedents of the SDSS BLQSOs population 
at lower redshifts. Note that this claim of a detection of a copiously accreting and obscured population of 
black holes at $z>6$ on X-ray image stacking by Treister et al. (2011) as well as its significance have been
disputed by Cowie et al. (2011) and Willott (2011). The issue can only be settled conclusively with more high 
redshift data that will hopefully be forthcoming. As noted in our earlier work, in particularly, VN09, with more 
comprehensive observational redshift coverage, several key assumptions in the modeling approaches can be tested.

In this paper we focus our modeling efforts on four redshifts of interest ($z=1.25$; $z=2$; $z=3.25$; $z=4.25$; to match 
the redshift bins in K10). At each redshift our models provide us with a sample of all SMBHs present at that particular cosmic time, and of 
the SMBHs that are actively accreting. From the SMBH mass and its Eddington ratio, $f_{\rm Edd}$, we can 
derive their bolometric luminosity: $\log(L_{\rm bol}/{\rm erg}\,{\rm s}^{-1}=38.11+\log(M_{\rm BH}/\msun)+\log(f_{\rm Edd})$. We 
apply a bolometric correction of 4.3 (as in K10) and select only quasars that are more luminous than the minimum luminosity 
determined by the flux limit described in Richards et al. (2006), the parent sample of K10. The threshold bolometric
luminosities corresponding to each bin are as follows. At $z=1.25$, $\log(L_{\rm bol}/{\rm erg\,s^{-1}})$=44.7;  at $z=2$: $\log(L_{\rm bol}/{\rm erg\,s^{-1}})$=45.2;  
at $z=3.25$: $\log(L_{\rm bol}/{\rm erg\,s^{-1}})$ 45.5; at $z=4.25$: $\log(L_{\rm bol}/{\rm erg\,s^{-1}})$=45.8.  Finally, we assume that the fraction 
of unobscured  quasars is 20\%, based on La Franca et al. (2005). Here we do not explicitly apply the evolutionary model of La Franca 
et al. (2005), where the fraction of obscured quasars depends on both redshift and luminosity, as the redshift range we are interested is 
beyond that explored by La Franca et al. (2005), but we note that when we apply their evolutionary model to the redshift range 
$z=1-3$, adopting the bolometric corrections of Marconi et al. (2004), we obtain consistent results. Recently Fiore et al. (2012) have
also derived and published BHMFs for the entire active SMBH population as opposed to just the BLQSOs as done by K10.

\section{Black hole growth models}

To trace the assembly history of black holes with cosmic time, Monte-Carlo realizations are performed 
to derive the  merger histories of dark matter haloes. We also trace the formation and growth of 
embedded black holes as a function of cosmic time as outlined below. 

\subsection{Black hole seeds}

To track the mass assembly history of black holes in the universe, we need to start with
seeds at high redshift. In the standard picture, the assumption is that the remnants of
the massive first stars (Pop III stars) provide the earliest seeds in the range of $50 - 100\,\msun$. 
We note here that whether the first stars were indeed this massive has been
called to question from the latest round of recent higher resolution simulation results 
where fragmentation occurs ubiquitously (Turk et al. 2009; Greif et al. 2011; Davis et al. 2011).
An alternate model for the  formation of massive seeds from the direct collapse of 
pre-galactic disks is presented in Lodato \& Natarajan (2007; 2006). In these models,
  there is a limited mass range of halos with a further narrow range in spins that are able 
  to form  seeds. However, contrary to the Pop III case, massive seeds with $M\approx 10^5-10^6M_{\odot}$ can 
  form at high redshift ($z>15$), when the intergalactic medium has not been significantly enriched by metals
\citep{Koushiappas2004,BVR2006,LN2006,LN2007}.  More details of this seeding model can be 
found in \citet{LN2006, LN2007}, wherein the development of non-axisymmetric spiral structures drives 
mass infall and accumulation in a pre-galactic disc with primordial
composition. The central mass accumulation that provides an upper limit to the SMBH seed mass that can form is 
given by:
\begin{equation}
M_{\rm BH}= m_{\rm d}M_{\rm halo}\left[1-\sqrt{\frac{8\lambda}{m_{\rm d}Q_{\rm c}}\left(\frac{j_{\rm d}}{m_{\rm d}}\right)\left(\frac{T_{\rm gas}}{T_{\rm vir}}\right)^{1/2}}\right] 
\label{mbh}
\end{equation}
for 
\begin{equation}
\lambda<\lambda_{\rm max}=m_{\rm d}Q_{\rm c}/8(m_{\rm d}/j_{\rm d}) (T_{\rm
  vir}/T_{\rm gas})^{1/2}
\label{lambdamax} 
\end{equation}
and $M_{\rm BH}=0$ otherwise. Here $\lambda_{\rm max}$ is the maximum
halo spin parameter for which the disc is gravitationally unstable,
$m_d$ is the gas fraction that participates in the infall and $Q_{\rm
c}$ is the Toomre parameter.  The efficiency of SMBH formation is
strongly dependent on the Toomre parameter $Q_{\rm c}$, which sets the
frequency of formation, and consequently the number density of SMBH
seeds. Guided by our earlier investigation, we set $Q_{\rm c}=2$ (the
intermediate efficiency massive seed model) as described in VLN08.

The efficiency of the seed assembly process ceases at large halo
masses, where the disc undergoes fragmentation instead. This occurs
when the virial temperature exceeds a critical value $T_{\rm max}$,
given by:
\begin{equation}
\frac{T_{\rm max}}{T_{\rm gas}}=\left(\frac{4\alpha_{\rm c}}{m_{\rm
d}}\frac{1}{1+M_{\rm BH}/m_{\rm d}M_{\rm halo}}\right)^{2/3},
\label{frag}
\end{equation}
where $\alpha_{\rm c}\approx 0.06$ is a dimensionless parameter measuring the
critical gravitational torque above which the disc fragments \citep{RLA05}.

To summarize the seeding model, every dark matter halo is characterized by its mass $M$
(or virial temperature $T_{\rm vir}$) and by its spin parameter
$\lambda$. The gas has a temperature $T_{\rm gas} = 5000$K. If
$\lambda<\lambda_{\rm max}$ (see eqn.~2) and $T_{\rm
vir}<T_{\rm max}$ (eqn.~\ref{frag}), then we assume that a seed BH of
mass $M_{\rm BH}$ given by eqn.~(\ref{mbh}) forms in the center. The
remaining relevant parameters are $m_{\rm d}=j_{\rm d}=0.05$,
$\alpha_{\rm c}=0.06$ and here we consider the $Q_{\rm c}=2$ case.
In the massive seed model, SMBHs form (i) only in haloes within a narrow range of virial
temperatures ($10^4$ K$ < T_{vir}<1.4\times10^4$ K), hence, halo
velocity dispersion ($\sigma \simeq 15\,{\rm km\,s}^{-1} $), and (ii)
for a given virial temperature all seed masses below $m_d\,M$ modulo the
spin parameter of the halo are allowed (see eqns.~1 and 3). The
seed MF peaks at $10^5\msun$, with a steep drop at
$3\times10^6\msun$.  We refer the reader to Lodato \& Natarajan (2007)
and VLN08 for a discussion of the mass
function (and related plots).  Here we stress that given points (i) and (ii) above, 
the initial seeds do not satisfy the local $M_{\rm BH} -\sigma$ relation, in fact the seed 
masses are not correlated with $\sigma$, rather they are correlated with the spin 
parameter of the dark matter halo.

In this paper, we once again contrast this model with a popular scenario that advocates the first black holes are the remnants of zero metallicity stars.  We assume that one Pop III star forms in metal-free halos with $T_{\rm vir}> 2000$ K \citep{Yoshida2006}. We assume a logarithmically flat initial MF, $dN/d\log M=$const, between $10\msun$ and $600 \msun$, where the upper limit comes from  \cite{OmukaiPalla}, and  suppose that seed black holes form when the progenitor star is in the mass range $40-140 \msun$ or $260-600\msun$ \citep{Fryer2001}. The remnant mass is  taken to be one-half the mass of the star.  

Both scenarios for black hole seed formation rely on zero metallicity gas. We model the evolution of metallicity by the `high feedback, best guess' model of \cite{Scannapieco2003}.  
\cite{Scannapieco2003} model metal enrichment via pair-instability supernovae winds, by following the expansion of spherical outflows into the Hubble flow. They compute the co-moving radius, at a given redshift, of an outflow from a population of supernovae that exploded at an earlier time.  Using a modification of the Press--Schechter technique, \cite{Scannapieco2002} compute the bivariate MF of two halos of arbitrary mass and collapse redshift, initially separated by a given co-moving distance. From this function they calculate the number density of supernovae host halos at a given co-moving distance from a `recipient' halo of  a given mass $M_h$ that form at a given redshift $z$. By integrating over this function, one can calculate the probability that a halo of mass $M_h$ forms from metal--free gas at a redshift $z$. When a halo forms in our merger tree we calculate the probability that it is metal-free (hence, it can form Pop III stars) and determine if this condition is satisfied. 

Every halo entering the merger tree is assigned a spin
parameter drawn from the lognormal distribution in $\lambda_{\rm
spin}$ found in numerical simulations, with mean $\bar \lambda_{\rm
spin}=0.05$ and standard deviation $\sigma_\lambda=0.5$ (Davis \& Natarajan 2009). We assume that the spin
parameter of a halo is not modified by its merger history, as no
consensus exists on this issue at the present time.

\subsection{Black hole growth}

We evolve the population of SMBH seeds according to simple models of self-regulation with the host. The main features of the 
models have been discussed elsewhere \citep{VN2009}.  We summarize below all the relevant modeling assumptions.  SMBHs in 
galaxies undergoing a major merger (i.e., having a mass ratio $>1:10$) accrete mass and become active. Each SMBH accretes 
an amount of mass, $\Delta M=9\times 10^7(\sigma/200\kms)^{4.24}\msun$, where $\sigma$ is the velocity dispersion after the merger.  
This relationship scales with the $M_{\rm BH}-\sigma$ relation, as it is seen today \citep{Gultekin2009}:
\beq
M_{\rm BH}=1.3\times10^8  \left(\frac{\sigma}{200 \kms} \right)^{4.24} \msun;
\eeq
the normalization in $\Delta M$ was chosen to take into account the contribution of SMBH-SMBH  mergers, without exceeding 
the mass given by the $M_{\rm BH}-\sigma$ relation.

We link the correlation between the black hole mass  and the central stellar velocity dispersion of the 
host with the empirical  correlation between  the central stellar velocity dispersion and the asymptotic circular 
velocity as $\sigma=v_{\rm c}/\sqrt[]{2}$) of galaxies \citep{Ferrarese2002; Pizzella2005; Baes2003}.
The latter is a measure of the total mass of the dark matter halo of the host galaxy. We calculate the circular velocity from the mass of the host halo and its redshift. 

The rate at which mass is accreted scales with the Eddington rate for the SMBH, and we set  either a fixed Eddington ratio 
of $f_{\rm Edd}=1$ (for Pop III seeds), $f_{\rm Edd}=0.3$ (for massive seeds), or an accretion rate derived from the distribution 
derived by \cite{Merloni08} (we apply this model to massive seeds only). The empirical distribution of Eddington ratios derived 
by Merloni \& Heinz (2008, MH08 thereafter) is fit by a function in $\log(L_{\rm bol}/L_{\rm Edd})$. The fitting function of 
the Eddington ratio distribution as a function of SMBH mass and redshift, is computed in 10 redshift intervals (from $z=0$ to $z=5$) for 
4 different mass bins ($6 < \log(M_{\rm BH}/\msun)< 7$; $7 < \log(M_{\rm BH}/\msun) < 8$; $8 < \log(M_{\rm BH}/\msun) < 9$; 
$9 < \log(M_{\rm BH}/\msun) < 10$), and then fit with an analytic function which is the sum of a Schechter function and a log-normal 
(A. Merloni, private communication). The  Eddington ratio distributions are then normalized to unity at every given mass and redshift.  
We dub the three models {\bf PopIII-Edd},  {\bf Massive-MH} and {\bf Massive-subEdd} respectively. Note that in the model names the first part refers to the type of seed and the second part refers to the kind of accretion history assumed. Therefore the model {\bf PopIII-Edd} refers to: initial sees from Pop III remnants always accreting at the Eddington rate; model {\bf Massive-MH} refers to initial massive seeds accreting with Eddington ratios drawn from the MH08 
distribution and the model {\bf Massive-subEdd}: initial massive seeds accreting $0.3X$ Eddington at all times.

 In the {\bf Massive-MH} model the accretion rate is not limited to the Eddington rate and mildly 
super-Eddington accretion rates (up to $f_{\rm Edd}\sim 10$) are possible and allowed as per MH08. 

For all three scenarios considered here, accretion starts after a dynamical timescale at the virial radius, 
$t_{\rm dyn}=10^8 {\rm yr}\; (R_{\rm vir}/100\, {\rm kpc})(v_{\rm vir}/100 \kms)^{-1}$,
and lasts until the SMBH, of initial mass $M_{\rm in}$, has accreted $\Delta M$.  The lifetime of an AGN therefore depends on how much mass it accretes during each episode:
\begin{equation}
t_{\rm AGN}=\frac{t_{\rm Edd}}{f_{\rm Edd}} \frac{\epsilon}{1-\epsilon}\,\ln({\frac {M_{\rm fin}}{M_{\rm in}}}),
\label{tagn}
\end{equation}
where $\epsilon$ is the radiative efficiency ($\epsilon \simeq 0.1$),  $t_{\rm Edd}=0.45$ Gyr and $M_{\rm fin}=\min[(M_{\rm in}+\Delta M),1.3\times10^8\,(\sigma/200 \kms)^{4.24}\msun]$.

We further assume that, when two galaxies hosting SMBHs merge, the SMBHs themselves merge within the merger 
timescale of the host halos, which is a plausible assumption for SMBH binaries formed after gas-rich galaxy mergers 
\cite{Dotti2007}. We adopt the relations suggested by \cite{Taffoni2003} for the merger timescale. Black holes are allowed 
to accrete during the merging process if the timescale for accretion, corresponding to the sum of the dynamical timescale 
and $t_{\rm AGN}$, is longer than the merger timescale. 

As outlined earlier, in propagating the seeds it is assumed that
accretion episodes and therefore growth spurts are triggered only by
major mergers. We find that in a merger-driven scenario for SMBH growth
the most biased galaxies at every epoch host the most massive SMBHs
that are most likely already sitting on the $M_{\rm BH} -\sigma$
relation.  Lower mass SMBHs (below $10^6\,\msun$) are instead off the
relation at $z = 4$ and even at $z = 2$. These baseline results are
{\bf independent of the seeding mechanism}.  
In the initial massive seeds scenario, most of the SMBH seeds start out {\bf
well above} the $z=0$ $\msigma$ relation, that is, they are `over massive'
compared to the local relation.  Seeds form only in haloes within a
narrow range of velocity dispersion ($\sigma \simeq 15\,{\rm
km\,s}^{-1}$ at the earliest epochs, see eqns.~1 and 3.  The SMBH mass
corresponding to $\sigma \simeq 15\,{\rm km\,s}^{-1}$, according to
the local $M_{\rm BH} -\sigma$ relation, would be $\sim 3\times 10^3
\msun$.  The MF instead peaks at $10^5\msun$ (Lodato \&
Natarajan 2007).  As time elapses, all haloes are bound to grow in
mass by mergers.  The lowest mass haloes, though, experience mostly
minor mergers, that do not trigger accretion episodes, and hence do
not grow the SMBH. The evolution of these systems can be described by
a shift towards the right of the $\msigma$ relation: $\sigma$
increases, but $M_{\rm BH}$ stays roughly constant. 

\begin{figure}
\includegraphics[width= \columnwidth]{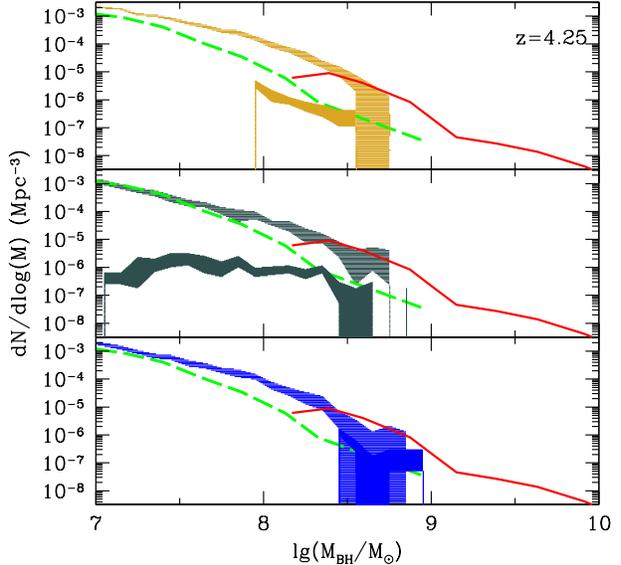}
\caption{The derived MF of SMBHs at $z=4.25$. The upper shaded curve in all three panels is the MF for all SMBHs including active and inactive ones. 
The lower (darker) shaded curve in all three panels is the MF for SMBHs that can be identified as BLQSOs. The dashed curve in all three panels
 is the MF of all SMBHs from MH08. The solid curve is the MF of BLQSOs from K10. The three panels refer to the models: ${\bf PopIII-Edd}$ 
 (uppermost panel), ${\bf Massive-MH}$ (middle panel), ${\bf Massive-subEdd}$ (bottom panel).}
\end{figure}

\begin{figure}
\includegraphics[width= \columnwidth]{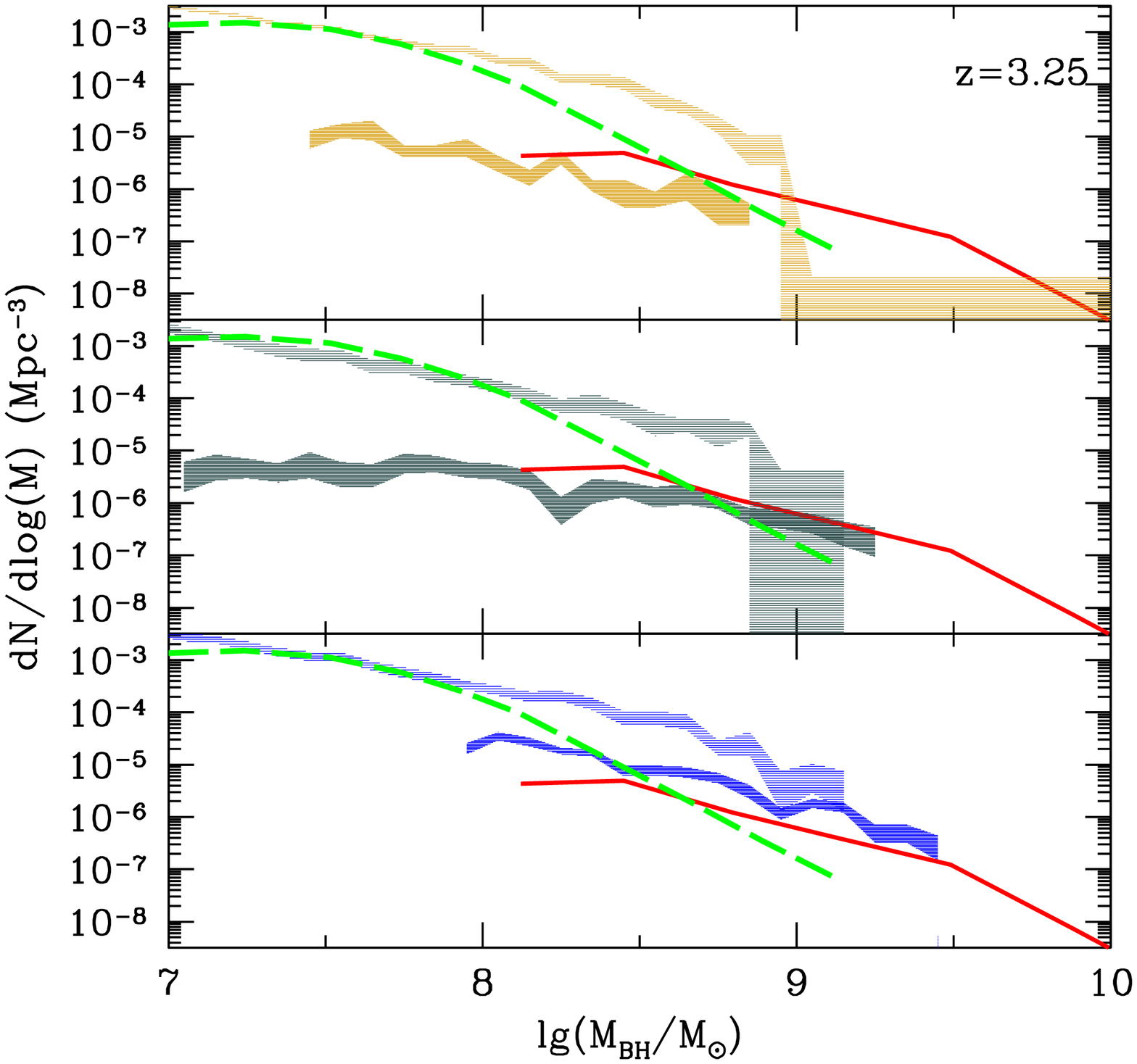}
\caption{The derived MF of SMBHs at $z=3.25$. The upper shaded curve in all three panels is the MF for all SMBHs including active and inactive ones. 
The lower (darker) shaded curve in all three panels is the MF for SMBHs that can be identified as BLQSOs. The dashed curve in all three panels
 is the MF of all SMBHs from MH08. The solid curve is the MF of BLQSOs from K10. The three panels refer to the models: ${\bf PopIII-Edd}$ 
 (uppermost panel), ${\bf Massive-MH}$ (middle panel), ${\bf Massive-subEdd}$ (bottom panel).}
\end{figure}

\begin{figure}
\includegraphics[width= \columnwidth]{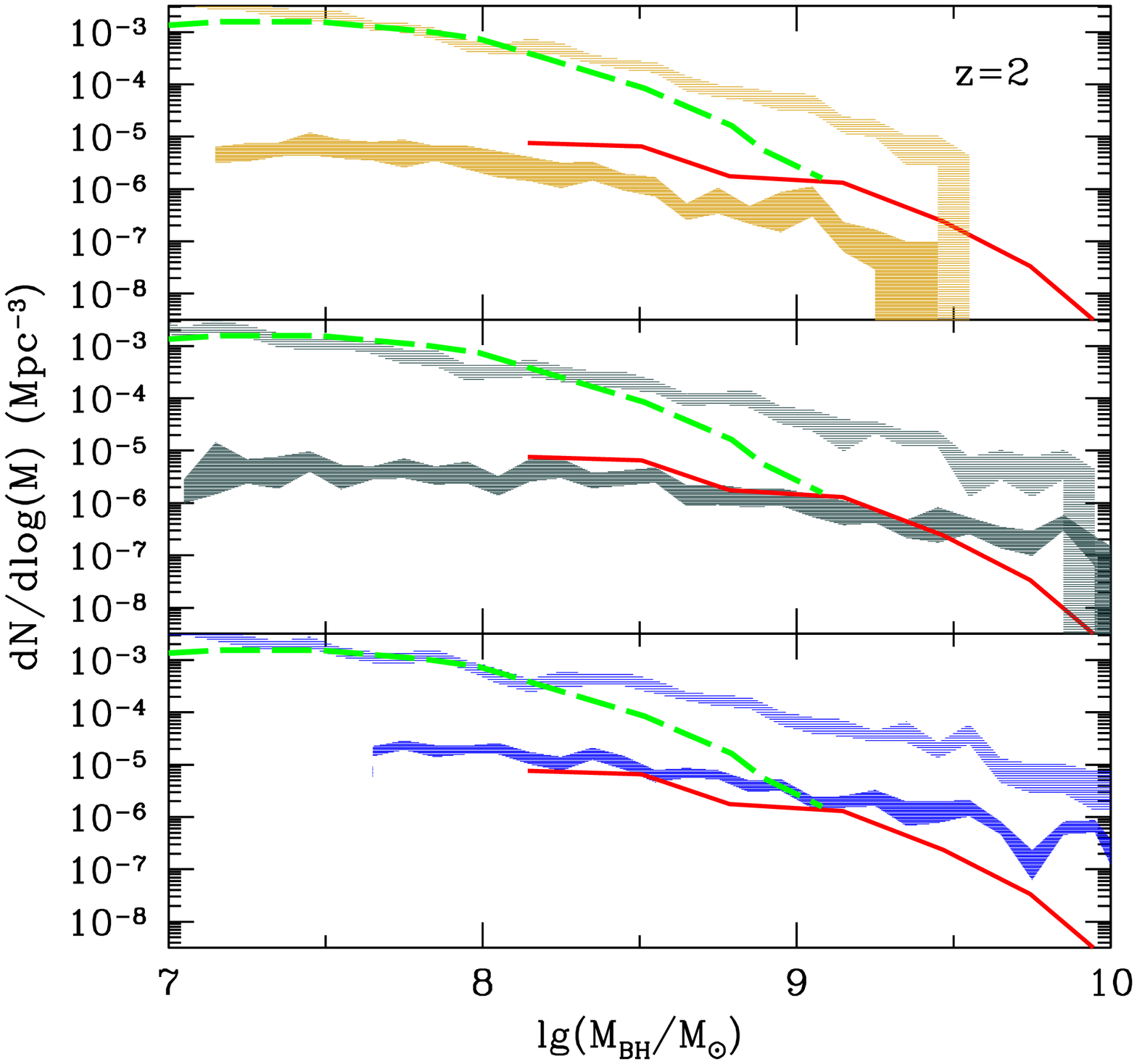}
\caption{The derived MF of SMBHs at $z=2$. The upper shaded curve in all three panels is the MF for all SMBHs including active and inactive ones. 
The lower (darker) shaded curve in all three panels is the MF for SMBHs that can be identified as BLQSOs. The dashed curve in all three panels
 is the MF of all SMBHs from MH08. The solid curve is the MF of BLQSOs from K10. The three panels refer to the models: ${\bf PopIII-Edd}$ 
 (uppermost panel),  ${\bf Massive-MH}$ (middle panel), ${\bf Massive-subEdd}$ (bottom panel).}
\end{figure}

\begin{figure}
\includegraphics[width= \columnwidth]{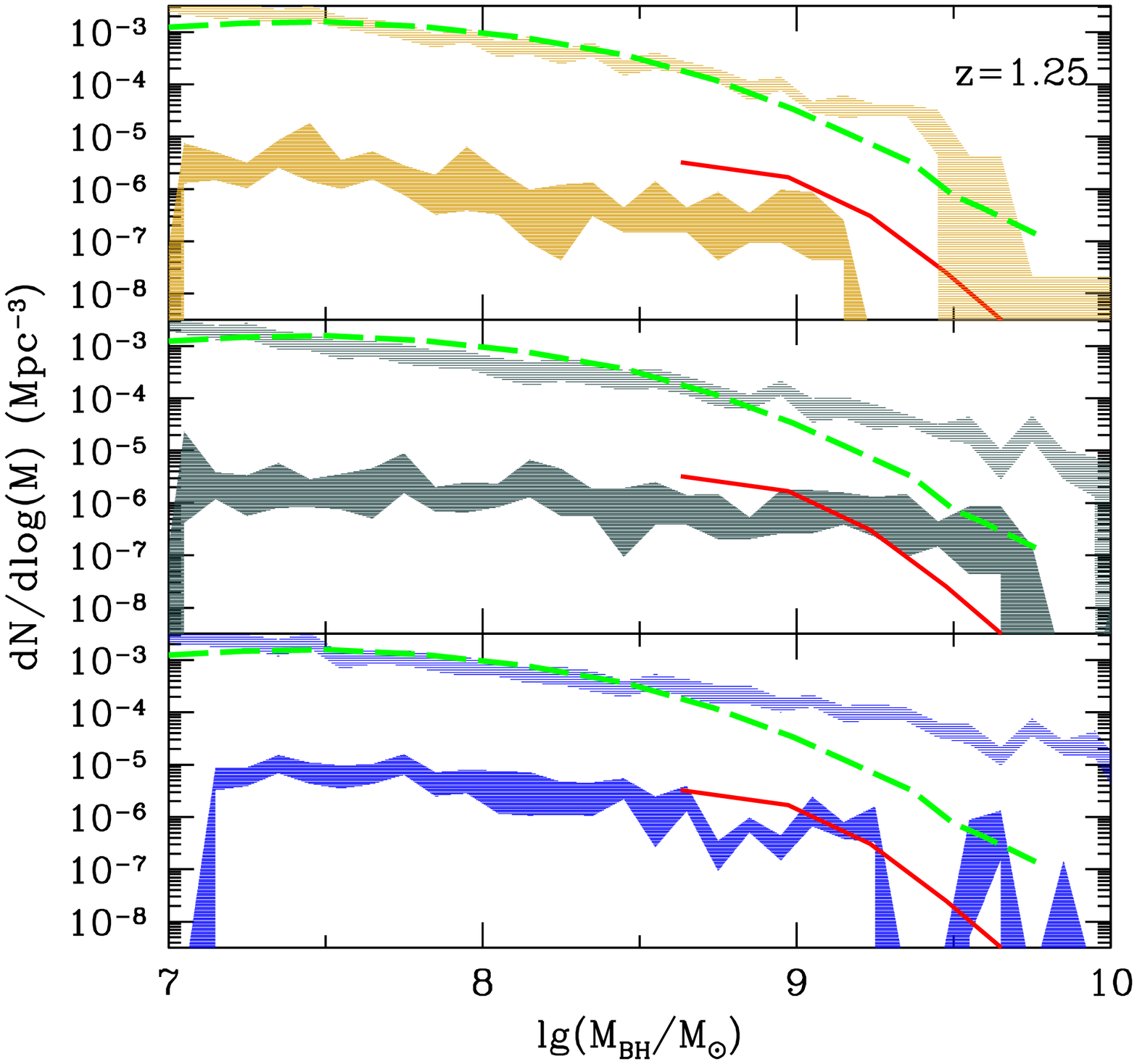}
\caption{The derived MF of SMBHs at $z=1.25$. The upper shaded curve in all three panels is the MF for all SMBHs including active and inactive ones. 
The lower (darker) shaded curve in all three panels is the MF for SMBHs that can be identified as BLQSOs. The dashed curve in all three panels
 is the MF of all SMBHs from MH08. The solid curve is the MF of BLQSOs from K10. The three panels refer to the models: ${\bf PopIII-Edd}$ (uppermost panel), ${\bf Massive-MH}$ (middle panel), ${\bf Massive-subEdd}$ (bottom panel).}
\end{figure}

\section{Results and Conclusions}

In what follows, we present a detailed comparison of the data with our three models, namely the {\bf PopIII-Edd}: initial seeds from Pop III remnants with accretion assumed at 
all times at the Eddington rate; {\bf Massive-MH}: wherein the initial massive seeds have accretion rates drawn from the distribution determined by MH08; {\bf Massive-subEdd}: initial massive seeds with accretion assumed at all times to be at $0.3 \times$ the Eddington rate. For all three models, we compare our derived MF of BLQSOs with that estimated by K10 and the MF of all SMBHs (the entire population that includes actively accreting + inactive black holes) with the MF predicted by MH08. Although the latter is also a theoretical MF, it is derived under completely different assumptions, as the MH08 model starts with SMBHs today and evolves them back in time via a continuity equation. We therefore consider this an interesting case of model confronting model (see also Volonteri \& Begelman 2010). 

\begin{figure}
\includegraphics[width= \columnwidth]{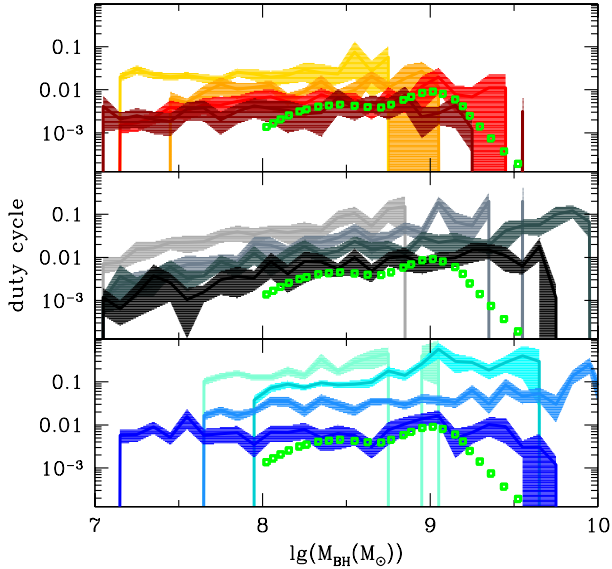}
\caption{The duty cycle of BLQSOs as a function of SMBH mass. The models shown here are again in the same order as before: 
${\bf PopIII-Edd}$ (uppermost panel), ${\bf Massive-MH}$ (middle panel), ${\bf Massive-subEdd}$ (bottom panel). In each panel, we show the duty cycle 
at $z=4$, $z=3$, $z=2$, and $z=1$ (from lightest to darkest shade respectively). The square symbols show the lower bound on 
the duty cycle at $z=1$ from K10.}
\label{dc}
\end{figure}

We compare our derived MFs for BLQSOs to K10 and our MFs for the entire BH population (active + inactive) to MH08 in Figures~1-4. We find the following:
\begin{itemize}

\item In the higher redshift slices at $z=4.25$ and $z=3.25$, independent of our models the MF derived in MH08 for all SMBHs is incompatible with that for the subset of BLQSOs from K10. In these instances the abundance of BLQSOs from K10 is in excess of that of all SMBHs from MH08 models. It is only in the lowest redshift bin at $z=1.25$ that the MFs of BLQSOs from K10 is lower than that of all SMBHs derived from the MH08 models. It is therefore clear that the Eddington ratio distributions derived from K10 and from  MH08 are most likely discrepant. 

%and we see this clearly in Figure 5.

\item At the highest redshift slice $z=4.25$, the light seed model {\bf PopIII-Edd}) shown in the top panel fails to grow SMBHs massive enough (more massive than $\sim 10^{8.5} M_{\odot}$) to populate the high-mass end of the MF of BLQSOs of K10. The discrepancy is roughly two orders of magnitude as clearly seen in the upper panel of Figure~1. Since the Eddington ratio was assumed to be unity to evolve this model, a deficit at the high mass end cannot be compensated for by altering the accretion physics any further.  Alternatively, if SMBHs can become ``over massive" at higher z (i.e., if they climb well above the standard $M_{\rm BH}-\sigma$ relation) and have phases of super-Eddington accretion (Volonteri \& Rees 2005), then enough SMBHs at the high mass end might be produced. We note that assuming the MH08 accretion model in conjunction with Pop III remnants (this model is not plotted in our figures) makes the discrepancy at the high mass end much worse, as the accretion rate for low mass black holes ($<1000\, \msun$) in the MH08 model peaks at sub-Eddington rates at all redshifts.

\item Both the massive seed models ({\bf Massive-subEdd} and {\bf Massive-MH}) also under-produce the K10 MFs at the highest redshift slice as seen clearly by looking at the middle and bottom panels in $z = 4.25$;

\item The MF of all existing SMBHs (active + inactive) from the {\bf PopIII-Edd}, {\bf Massive-subEdd} and {\bf Massive-MH} models 
overestimates MH08's MF for all SMBHs at all masses $>3\times 10^7\, \msun$ at $z=3.25$, while there is reasonably
good agreement with K10 for the BLQSO MF for the {\bf Massive-MH} model. 

\item At lower redshifts ($z=1.25$ and $z=2$) both massive seed models start to overestimate the MF of BLQSOs  by K10 at $M_{\rm BH}>10^9\,\msun$.  This is not unexpected, as we have not implemented a cut-off to mimic the depletion of available gas in massive galaxies. 

\item All massive seeds models ({\bf Massive-subEdd} and {\bf Massive-MH}) overestimate the MF of all SMBHs (active + inactive) from MH08 at $M_{\rm BH}>10^{8.5}\, \msun$. At $z<2$ we  cannot clearly assess whether MH08 might underestimate the total MF. We are, however, inclined to attribute the discrepancy to our models,  for the same reason that we likely overestimate the MF of quasars as we do not take into account changes in the gas inventory at late times in the universe. The {\bf PopIII-Edd} model gets instead into better alignment at lower and lower redshifts. At $z = 2$ these MFs begin to agree up to $10^8\,M_{\odot}$ and at $z = 1.25$ they fall into excellent agreement up to $10^9\,M_{\odot}$.

\item For model {\bf Massive-MH} we can estimate the typical Eddington ratio of SMBHs powering BLQSOs. We find that most luminous quasars at the highest redshift have accretion rates close to or even slightly larger than the Eddington rate (the fraction of super-Eddington accreting SMBHs is $\simeq$ 60\% at $z=3$; $\simeq$ 40\% at $z=2$; $\simeq$ 20\% at $z=1$). At lower redshift, the flux limit of the SDSS survey corresponds to lower luminosities, and less powerful accretors are selected. In general, we find that the accretion rates in the models overestimate the accretion rates derived by K10.

\item We estimate the duty cycle, defined as the fraction of SMBHs that are active and detectable as BLQSOs, and show the results for the three models ({\bf PopIII-Edd}, {\bf Massive-subEdd}, {\bf Massive-MH}) in Figure~\ref{dc} at $z=1$, $z=2$, $z=3$, $z=4$ (dark to light shade of colours). Comparing the theoretical duty cycle with the duty cycle estimated by K10 at $z=1$, we find that model {\bf PopIII-Edd} is consistent at all masses; while models {\bf Massive-MH} and {\bf Massive-Edd} overestimate the duty cycle at $M_{\rm BH}>10^9\, \msun$ (as expected from the MF shown in Figures 3 and 4.). 

\item We find that the duty cycle increases with increasing redshift for all our models, although more strongly for the {\bf Massive-subEdd} model. This is due to the fact that this model is the one with the lowest average accretion rate ($f_{\rm Edd}=0.3$, versus  $f_{\rm Edd}=1$ for the {\bf PopIII-Edd} and  $f_{\rm Edd}\gsim 1$ in {\bf Massive-MH} at $z>3$, therefore, from Equation~\ref{tagn} the BHs in the {\bf Massive-subEdd} model need to accrete for a longer time in order to reach the final mass set by the $M_{\rm BH}-\sigma$ relation.

\end{itemize}

In conclusion, the implications for seeding models are that light seeds always under-produce the bright end
of the BLQSO LF for high black hole masses. This is a serious issue as they are assumed to be already accreting at the optimal
Eddington rate. 

Moreover, according to recent work (Turk et al. 2009; Greif et al. 2011; Davis et al. 2011) it appears now that (i) the IMF of the 
first stars may not be biased high as fragmentation might occur during the formation process and (ii) due to turbulence the formation 
of the first stars might get delayed. So the {\bf PopIII-Edd} model aside from being unable to match the black hole MF derived 
from BLQSOs from the SDSS as shown here, might in fact not be the most efficient channel to produce
seeds. It is clear that a channel to form more massive black hole seeds at high redshifts is required to
start matching observations of the bright end (massive end of the black hole MF) from the
earliest to intermediate redshifts ($z \sim 1$), and not only at the highest redshifts as previously noted 
(e.g., Haiman 2004; Yoo \& Miralda-Escud{\'e} 2004; Shapiro 2005;  Volonteri \& Rees 2005, 2006). 
{\bf Therefore, low-mass SMBH seeds such as Pop III star remnants require exceptional growth conditions not only 
to explain the existence of rare objects such as $z>6$ quasars, but also to explain the unexceptional 
population of luminous quasars at $z\simeq 2-3$.} 

Over-predicting the BHMF at the high mass end is typically not a serious problem as lowering the Eddington rate would
fix that.  For instance, Natarajan \& Treister (2009) propose the existence of an upper mass limit to BHs at all epochs, arising from the shutting off of accretion due 
to self-regulation of the gas supply in the galactic nuclei. According to their estimate this upper limit is given by:
\begin{eqnarray}
M_{\rm BH}  \sim 5 \times 10^9\, ({\frac{\sigma}{350\,{\rm km\,s}^{-1}}})^5\,\msun.
\end{eqnarray}
K10 estimate an upper mass limit at $z \sim 1$ for a black hole hosting a BLQSO to be $M_{\rm BH} \sim 3 \times 10^{10}\,\msun$.
Utilizing the limit proposed by Natarajan \& Treister (2009), if a BH shuts off at the end of this BLQSO episode at $z \sim 1$, it is likely 
hosted in a galaxy with a velocity dispersion of $\sigma \sim$ 440 km/s, which corresponds to the typical velocity dispersion of the 
brightest central galaxy in a cluster. This fits in nicely with the view of down-sizing wherein accretion is noted to shut off at $z \sim 1$ in these BCGs in clusters. 

\section*{Acknowledgements}

PN acknowledges support from the John Simon Guggenheim Foundation and 
a residency at the Rockefeller Bellagio Center where a portion of this work was completed. She also thanks
the Institute for Theory and Computation at Harvard for hosting her during her Guggenheim Fellowship year.

\end{document}